\documentclass[]{spie}  

\usepackage{amsmath,amsfonts,amssymb}
\usepackage{graphicx}
\usepackage[colorlinks=true, allcolors=blue]{hyperref}

\title{XSLIDE (X-Ray Spectral Line IDentifier and Explorer): a quick-look tool for XRISM}

\author[a,b]{Efrem Braun}
\author[a,c]{Chris Baluta}
\author[a,d]{Trisha F. Doyle}
\author[a,d]{Patricia L. Hall}
\author[a,b]{Robert S. Hill}
\author[a]{Matthew P. Holland}
\author[a,e,f]{Michael Loewenstein}
\author[g]{Eric D. Miller}
\author[a,b]{Michael C. Witthoeft}
\author[a,f,h]{Tahir Yaqoob}
\affil[a]{NASA Goddard Space Flight Center, Greenbelt, Maryland, USA}
\affil[b]{ADNET Systems, Inc., 6720B Rockledge Drive, Bethesda, MD, USA}
\affil[c]{KBR, Inc., 601 Jefferson Street, Houston, TX, USA}
\affil[d]{INNOVIM, LLC, 6401 Golden Triangle Drive, Greenbelt, MD, USA}
\affil[e]{University of Maryland, Department of Astronomy, College Park, MD, USA}
\affil[f]{Center for Research and Exploration in Space Science and Technology, NASA Goddard Space Flight Center, Greenbelt, MD, USA}
\affil[g]{Massachusetts Institute of Technology, Kavli Institute for Astrophysics and Space Research, Cambridge, MA, USA}
\affil[h]{University of Maryland Baltimore County, Center for Space Sciences and Technology, Baltimore, MD, USA}

\authorinfo{Further author information: (Send correspondence to E.B.)\\E.B.: E-mail: efrem.braun@nasa.gov}

\begin{document} 
\maketitle

\begin{abstract}
We present XSLIDE (X-Ray Spectral Line IDentifier and Explorer), a graphical user interface that has been designed as a quick-look tool for the upcoming X-Ray Imaging and Spectroscopy Mission (XRISM). XSLIDE is a simple and user-friendly application that allows for the interactive plotting of spectra from XRISM's Resolve instrument without requiring the selection of models for forward-fitting. XSLIDE performs common tasks such as rebinning, continuum fitting, automatically detecting lines, assigning detected lines to known atomic transitions, spectral diagnostics, and more. It is expected that XSLIDE will allow XRISM’s scientific investigators to rapidly examine many spectra to find those which contain spectral lines of particular interest, and it will also allow astronomers from outside the field of high-resolution X-ray spectroscopy to easily interact with XRISM data.

\end{abstract}

\keywords{XRISM, X-ray Astronomy, Astronomical Software}

\section{INTRODUCTION}


The X-Ray Imaging and Spectroscopy Mission (XRISM) Resolve instrument is expected to collect high energy X-ray spectra of unprecedented quality\cite{Tashiro18}. Rigorous analyses of these spectra require iterative forward-fitting, in which a model spectrum is specified and convolved with the instrument response to produce predicted photon counts, which are compared with the observed photon counts, and followed by adjustment of the model parameters to minimize the difference between the predicted and observed counts\cite{Smith11}. This procedure can be overly complicated and time-consuming for one who wants to quickly examine a spectrum, as it requires the user to specify and then refine a model spectrum.

As a precursor to this rigorous analysis, the XRISM Science Data Center (SDC) has developed XSLIDE (X-Ray Spectral Line IDentifier and Explorer), which allows scientific investigators to rapidly browse and survey available Resolve spectra. This will allow for easy access to XRISM data for astronomers from outside the field of high-resolution X-ray spectroscopy, quick browsing/surveying of available XRISM data prior to a decision to focus on a given spectrum, and simple extraction and characterization of XRISM spectra for non-publishing use such as writing proposals or giving presentations.

\begin{figure} [ht!]
\begin{center}
\begin{tabular}{c}
\includegraphics[width=12.5cm]{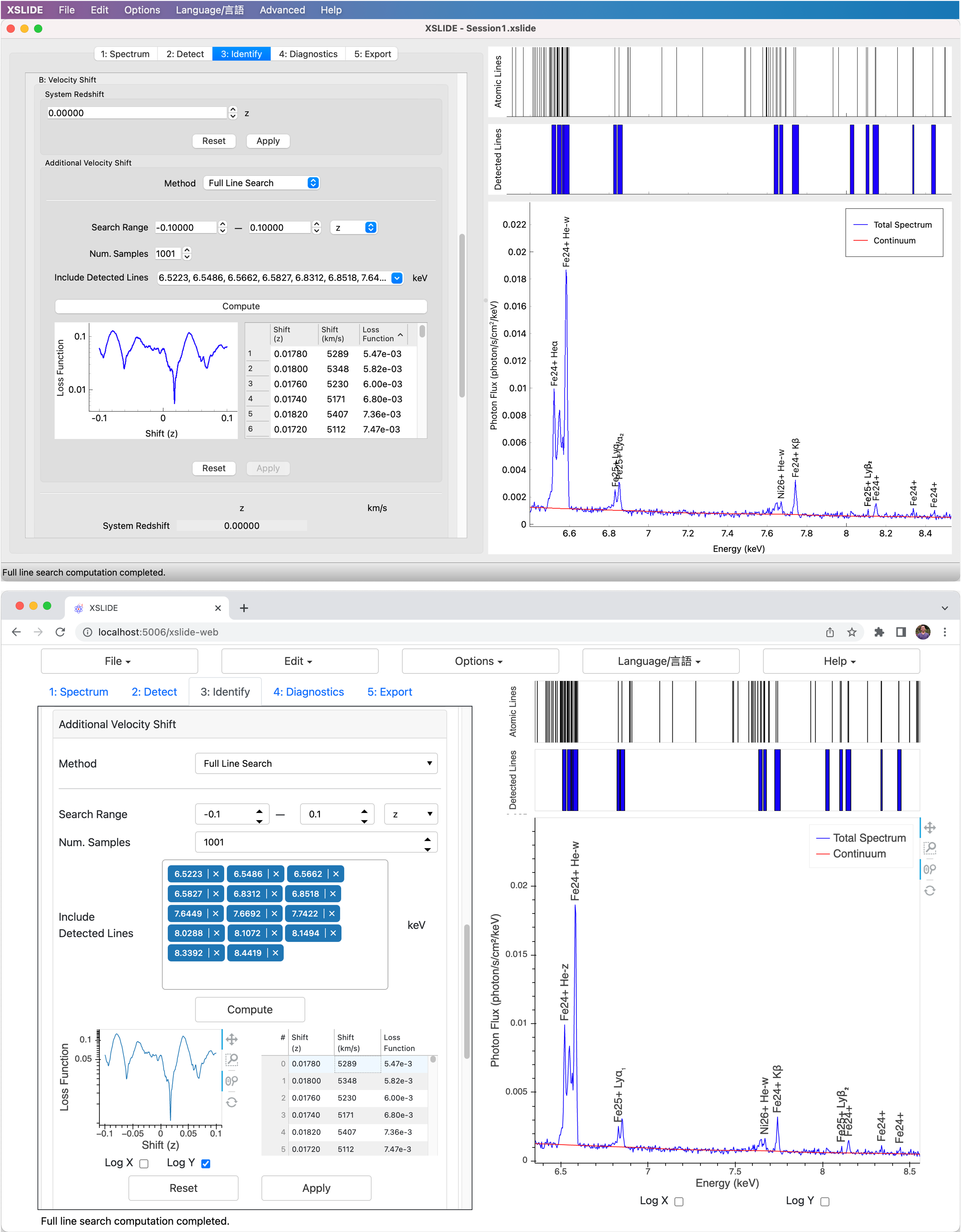}
\end{tabular}
\end{center}
\caption
{\label{fig:xslide-desktop-and-web}
Screenshots of the XSLIDE user interfaces, demonstrating the similarity between the desktop (above) and web (below) versions.}
\end{figure}

XSLIDE has been designed as a simple graphical user interface available as a desktop or web application (see Fig.~\ref{fig:xslide-desktop-and-web}) which allows users to interactively plot X-ray spectra. It walks the user through loading and modifying a spectrum, detecting lines, identifying those lines with known lines from atomic databases, performing diagnostics, and exporting the results. Following exploratory work performed with XSLIDE, more rigorous tools such as XSPEC\cite{Arnaud96} can be used to do further analyses on spectra of interest.

In this paper, we give a brief overview of XSLIDE. We discuss the methodology by which it allows users to view a spectrum with no selection of models for forward-fitting required and we assess the validity of the assumptions required for this process. We then outline the features of XSLIDE and show how the user is walked through the application step-by-step. Finally, we discuss the software architecture of XSLIDE to give an idea of how the program is built.

\section{Methodology}

A spectrometer obtains counts ($C$) within particular instrument channels ($I$). This is related to the source spectrum ($S$, in $\frac {\text{photon}} {\text{cm}^2 \text{ second keV}}$) by
\begin{equation}
\label{eq:specmaster}
C(I) = T \int R_{\text{RMF}}(I, E) R_{\text{ARF}}(E) S(E) dE \, ,
\end{equation}
where $T$ is the total observation time, $R_{\text{RMF}}$ is the unitless redistribution matrix file (RMF) which gives the probability of a photon of a given energy ending up as a count in a particular channel, and $R_{\text{ARF}}$ is the ancillary response file (ARF) which contains the effective area in $\frac {\text{cm}^2 \text{ count}} {\text{photon}}$\cite{Smith11}.

As Equation~(\ref{eq:specmaster}) cannot be analytically inverted to solve for $S$, it is typically solved by forward-fitting, whereby a model spectrum $S$ is specified, and the resulting predicted counts are compared to the observed counts, adjusting the parameters of the model to minimize some statistical measure of the error between the predicted counts and the observed
counts\cite{Smith11}. This procedure can be quite complicated and time-consuming as it requires the user to specify and then refine the model.

XSLIDE instead assumes that the RMF is a diagonal matrix (providing an ideal one-to-one mapping between incident photon energy and detector channel) and that the ARF is slowly varying such that it is approximately constant between neighboring instrument channels. This allows for $S$ to be discretely solved for as
\begin{equation}
\label{eq:specsolved}
   S(E) = \frac{C(I)}{R_{\text{ARF}}(E) T \Delta E } \, ,
\end{equation}
which can be computed using the ARF to map the instrument channel to the spectrum energy. This allows a spectrum to be viewed directly, requiring the user to provide only the spectrum and ARF files.

\begin{figure} [ht]
\begin{center}
\begin{tabular}{c}
\includegraphics[width=8cm]{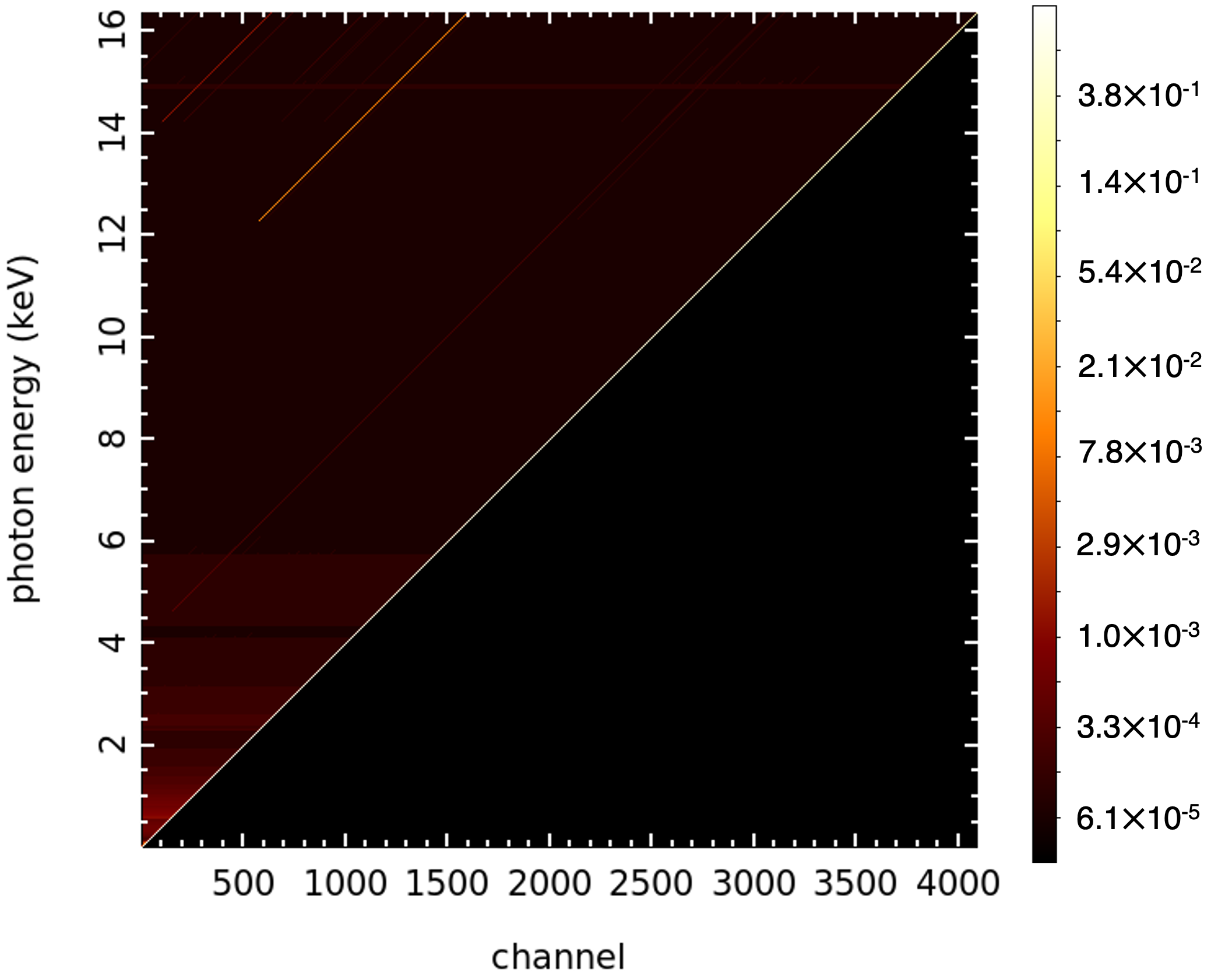}
\end{tabular}
\end{center}
\caption
{\label{fig:hitomi-rmf}
A sample RMF from the Hitomi SXS instrument, showing a sharp peak along the diagonal. The RMF shown was generated using the Hitomi ``sxsrmf'' tool for a single SXS pixel and for HighRes events, using the ``x-large'' format to include full input model resolution with all response components. For testing and display purposes, the level of the electron continuum component was increased by a factor of ten over the CalDB value.}
\end{figure}

\subsection{Validity}

As the RMF is not truly diagonal, the assumption of its being diagonal will result in an inexact spectrum; hence the need for the use of XSLIDE to be followed up with more rigorous tools such as XSPEC\cite{Arnaud96}. However, if the RMF is strongly peaked along the diagonal without significant off-diagonal contributions, then the locations of the spectral peaks will be correct, while only the peak widths will be incorrect. Thus, the approximate spectra gained using this assumption will be valid for the purpose of line identification and certain analyses. Although RMFs are not yet available for the XRISM mission, it is expected that the XRISM Resolve instrument's RMF will be similar to that of the Hitomi SXS instrument\cite{Tashiro18}, for which the RMF has been observed to be strongly peaked along the diagonal (see Fig.~\ref{fig:hitomi-rmf}).

The assumption of the ARF being locally constant can bring about spurious lines or hide lines at energies where the ARF changes quickly. The user can verify that the ARF is not changing quickly by overlaying it on the spectrum. In addition, if any detected lines are found in regions where the ARF is changing quickly, a warning is shown to the user that these lines may be artifacts.

\begin{figure} [ht]
\begin{center}
\begin{tabular}{c}
\includegraphics[width=13.5cm]{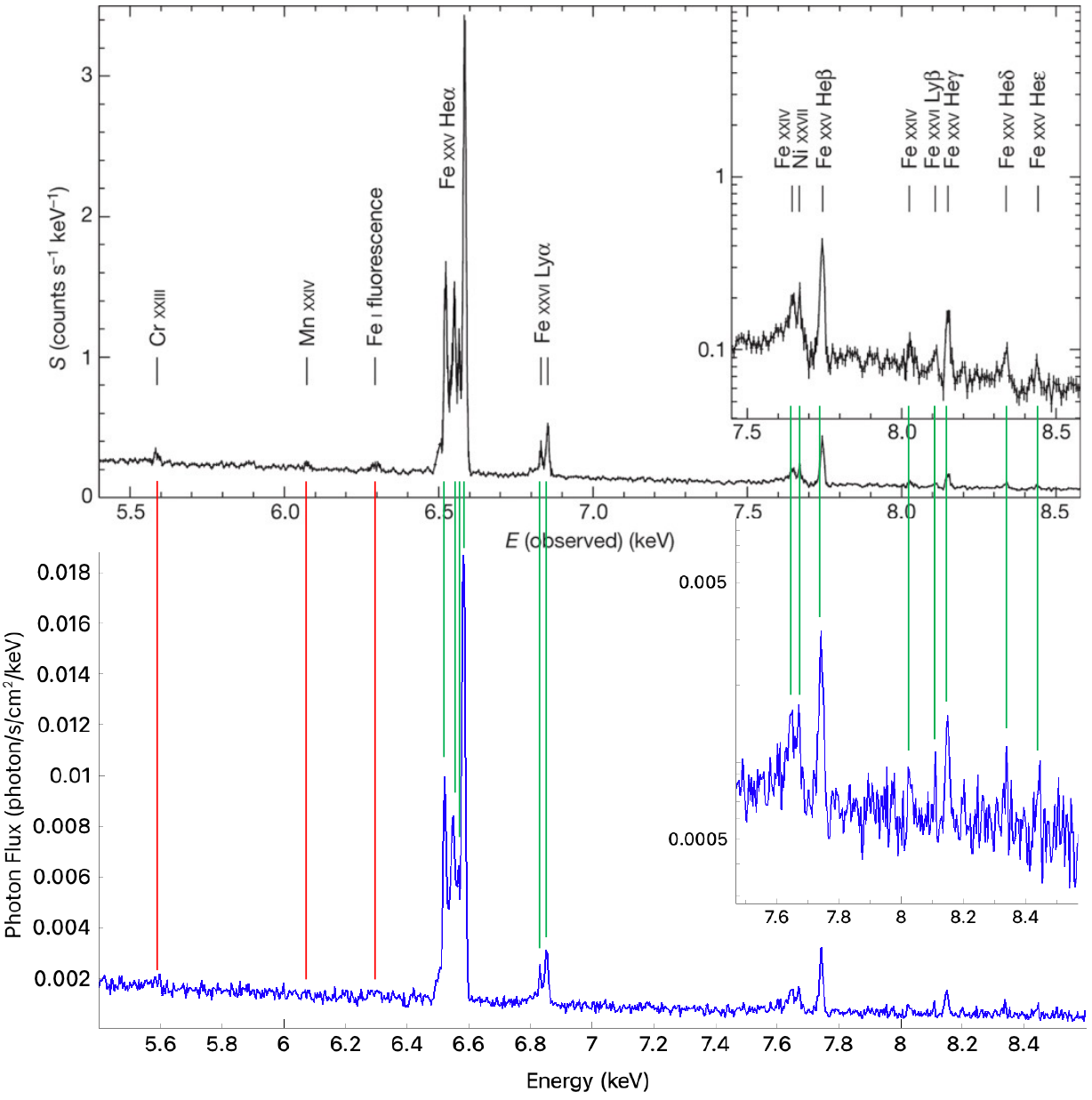}
\end{tabular}
\end{center}
\caption
{\label{fig:spectrum-comparison-perseus}
Comparison between spectra of the Perseus cluster taken by the Hitomi SXS instrument as shown in Tashiro et al.\cite{Hitomi16} (above) and as produced by XSLIDE (below). The XSLIDE image used the spectrum in ObsID 100040020, rebinned to 4.0 eV. See Tashiro et al.\cite{Hitomi16} for full details of their methods used. Note that some differences between the spectra are expected due to Tashiro et al.\cite{Hitomi16} having combined multiple observations for a longer cumulative exposure time and having performed additional corrections, which likely explain the small peaks at 5.6, 6.1, and 6.3 keV that were not observed with XSLIDE.}
\end{figure}

As a check on whether these assumptions are sufficiently valid for XSLIDE to be used reliably with spectra taken by the XRISM Resolve instrument, we compared the spectrum of the Perseus cluster taken by the Hitomi SXS instrument as produced rigorously in Tashiro et al.\cite{Hitomi16} with a spectrum generated by XSLIDE. As Fig.~\ref{fig:spectrum-comparison-perseus} shows, the spectrum produced by XSLIDE well-reproduces the more rigorously produced spectrum.

\section{Running XSLIDE}
XSLIDE is designed to be simple and easy to use. The user is guided through ordered steps and substeps, with little need for actions to be taken outside this guided procedure. In this section, we give a brief overview of these steps to illustrate XSLIDE's capabilities.

\subsection{Step 1: Spectrum}

In this step, the spectrum is loaded and modified.

The user can either load the spectrum and ARF files from the local desktop or perform an online query of NASA's High Energy Astrophysics Science Archive Research Center (HEASARC)\cite{heasarc}. After the spectrum is loaded, the user has the option to clip the edges of the spectrum, which can be used for such purposes as removing artifacts due to artificially low ARF. The user can then rebin the spectrum to improve the signal-to-noise ratio.

\begin{figure} [ht]
\begin{center}
\begin{tabular}{c}
\includegraphics[width=12.5cm]{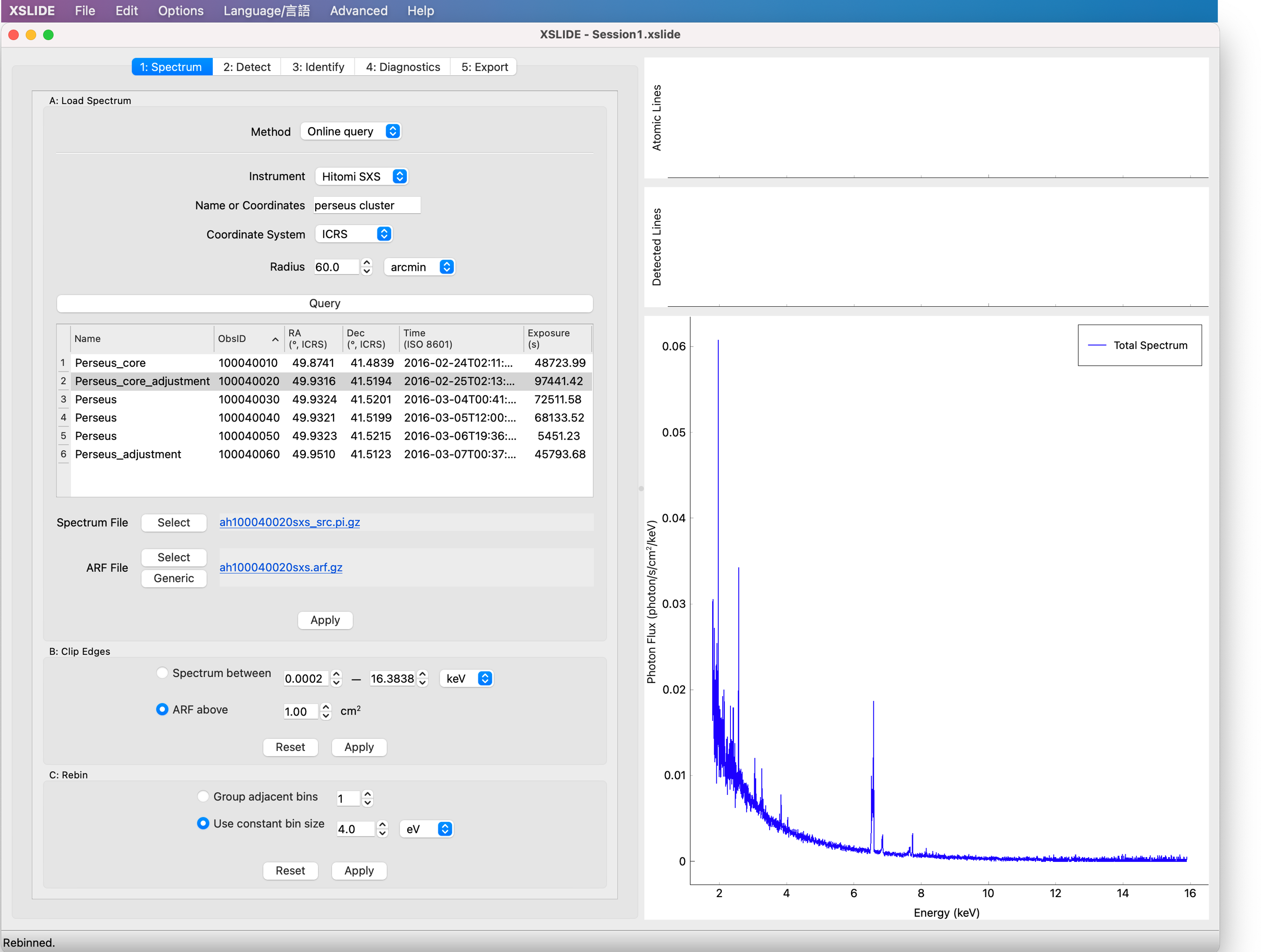}
\end{tabular}
\end{center}
\caption
{\label{fig:step1}
Screenshot of the desktop version of XSLIDE after Step 1 is complete.}
\end{figure}

Fig.~\ref{fig:step1} shows a sample screenshot of XSLIDE after Step 1 is complete.

\subsection{Step 2: Detect}

In this step, spectral lines are detected.

The first substep in accomplishing this is fitting a continuum, which can help with the next step of automatically detecting lines when the lines are placed over a non-flat continuum, such as can occur with bremsstrahlung radiation. Two methods exist for continuum fitting. First, XSLIDE can automatically fit a power law function in a piecewise manner to segments of the spectrum. Second, the user can manually specify points, which XSLIDE will then interpolate to form the continuum.

XSLIDE then provides two methods for automatic line detection. The piecewise statistics method works by first calculating the average and standard deviation statistics of piecewise segments of the spectrum, and then setting aside points with flux outside a user-specified number of standard deviations of the segment’s average as outliers. This process is then repeated iteratively several times, calculating the average and standard deviation statistics for the segments using only points that are not considered outliers until the statistics converge. Sets of consecutive outliers are considered detected lines. The second method works using the continuous wavelet transform (CWT), convolving the spectrum with Ricker wavelets of user-defined widths to effectively search the spectrum for regions which match the shape of the wavelets. The parts of the spectrum which better match the shape yield higher CWT coefficients, and points with local maximums in these CWT coefficients and which are above a user-defined signal-to-noise ratio are labeled as peaks. Each of the two methods has strengths and weaknesses. The piecewise statistics method can search for lines of varying widths, and it finds both emission and absorption lines, but it can fail to identify closely-spaced peaks as being separate lines if the valleys are sufficiently far away from the continuum, and it requires appropriate binning and continuum fitting to work. The CWT method is better at separately identifying closely-spaced peaks, and proper binning and continuum fitting are not critical to its success, but the user needs to specify the peak widths to search for, and it is mainly only capable of searching for either emission or absorption lines (but not both) in a given spectrum.

After automatically detecting lines, the user can then make manual adjustments as required. Undetected lines can be added, spurious lines can be removed, and existing lines can be narrowed or broadened as required.

We note that automatic line detection is far from a solved problem; neither of the methods provided in XSLIDE will be able to find all the peaks and discard all of the noise that can be readily discerned with a human eye. Rather, the point of XSLIDE's automatic line detection is to quickly find a spectrum's dominant features to determine if further analysis with a rigorous tool such as XSPEC\cite{Arnaud96} is warranted.


\subsection{Step 3: Identify}

In this step, the detected lines are identified using an atomic line database.

The user must first choose a set of atomic lines to use. XSLIDE comes packaged with AtomDB\cite{Smith01} version 3.0.9, several atomic databases custom-generated with XSTAR\cite{Mendoza21, Kallman21, Kallman01} version 2.58, and a neutral species database that was created using the data in Table 1-3 of the X-Ray Data Booklet\cite{Kortright09}. These built-in databases are expected to meet the requirements of the majority of XSLIDE users, and XSLIDE also allows users with niche needs to load their own atomic databases. Atomic line filters are implemented to allow the user to focus on particular ions or ``stronger'' lines with cutoffs for oscillator strength or emissivity.

Before the detected lines can be matched with the atomic lines, the spectrum’s velocity shift (generally referred to simply as “redshift”) needs to be accounted for so that the lines are energetically aligned. Several methods are available to specify the velocity shift. The manual method works by having the user simply enter a value. With the single line alignment method, the user selects a detected line and an atomic line, and the velocity shift gets set such that the energies of these two lines will be equal. This method of velocity shift determination is often done for intracluster medium spectra using the prominent Fe24+ line at 6.7 keV, for example. A more experimental method is the full line search, where a user-specified range of velocity shifts are generated, and for each velocity shift, a loss function is determined by calculating the deviation between each detected line and its nearest atomic line, summing the absolute values of these deviations. Velocity shifts with lower loss functions are worth further examination and can be applied by the user. This method is shown in Fig.~\ref{fig:xslide-desktop-and-web}, where it can be observed that a velocity shift of 0.0178 was found for the Perseus cluster observation by the Hitomi SXS instrument, quite close to the value of 0.01756 calculated in Tashiro et al.\cite{Hitomi16}.

Once the detected lines and atomic lines are aligned with each other, it is simple to identify the detected lines. XSLIDE simply matches each detected line with all atomic lines that are within the energy range determined by the edge points of the detected line. To help the user identify detected lines for which multiple atomic lines are possible, XSLIDE shows the oscillator strength and emissivities of the atomic lines to inform the user of the strength of the various lines. The user can choose to display labels next to the lines on the graph to indicate the identified atomic transitions.

Fig.~\ref{fig:xslide-desktop-and-web} shows sample screenshots of XSLIDE after Step 3 is complete.


\subsection{Step 4: Diagnostics}

In this step, the user views diagnostics that are computed using the identified lines. These diagnostics consist of ratios of line fluxes which have been found to correlate with system properties that may be of interest, such as temperature or ionization parameter.

The purpose of the XSLIDE diagnostics is only to provide a preliminary analysis, alerting the user to the possibility of the spectrum's diagnostic potential. The diagnostics should not be considered quantitatively reliable, as the detected line fluxes used for these calculations are only approximations which are highly dependent on the continuum and the number of points that define the detected lines. Any overlap between detected lines can drastically distort these calculated fluxes. To actually compute reliable diagnostics, a more rigorous tool such as XSPEC\cite{Arnaud96} is required.

\begin{figure} [ht]
\begin{center}
\begin{tabular}{c}
\includegraphics[width=14.5cm]{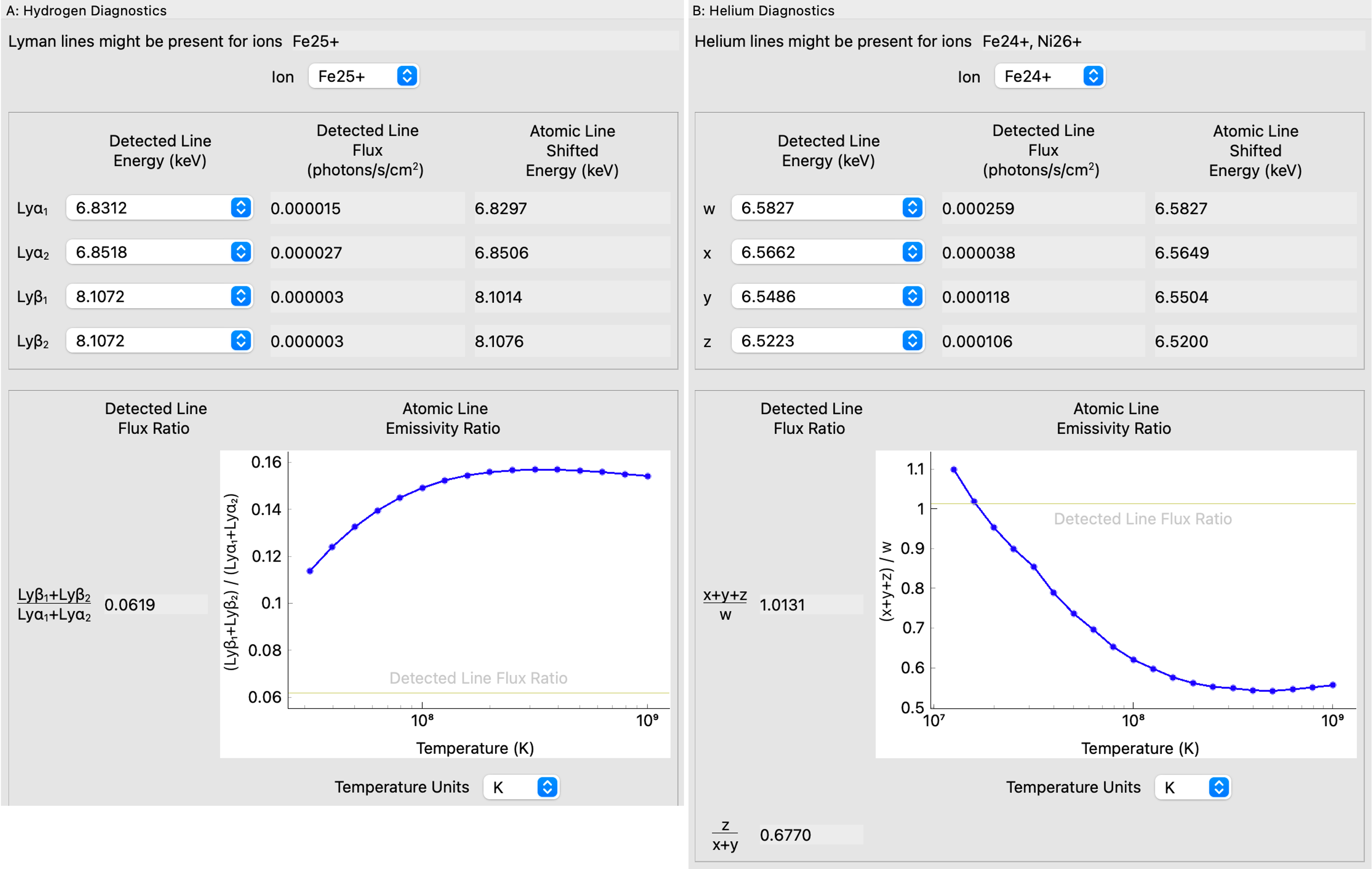}
\end{tabular}
\end{center}
\caption
{\label{fig:step4}
Partial screenshots of the hydrogen (left) and helium (right) diagnostics from Step 4 of XSLIDE. The hydrogen diagnostics shown indicate a temperature less than 30 million K (emissivities are not provided in AtomDB for temperatures low enough to give a Ly$\beta$/Ly$\alpha$ ratio of 0.0619 for Fe25+), and the helium diagnostics shown indicate a temperature of 16.2 million K.}
\end{figure}

Three diagnostics are available. In the hydrogen diagnostics (see Fig.~\ref{fig:step4} for a sample screenshot), ions for which the Lyman lines have been identified in the spectrum will have the Ly$\beta$/Ly$\alpha$ ratio calculated. This ratio will also be computed for the selected ion using the data in the atomic database as a function of temperature or ionization parameter and shown to the user in a graph; by seeing where the calculated ratio of the detected line fluxes intersect this graph, the temperature or ionization parameter may be approximated (as long as the assumptions that were used to create the selected atomic database are valid).

In the helium diagnostics (see Fig.~\ref{fig:step4}), ions for which the common helium-like lines (w, x, y, and z; also known as resonance (R), intercombination (I), intercombination (I), and forbidden (F)) have been identified in the spectrum will have the G ($\frac{x+y+z}{w}$) and R ($\frac{z}{x+y}$) ratios calculated. Similarly to the hydrogen diagnostics, the G ratio will also be computed for the selected ion using the data in the atomic database as a function of temperature or ionization parameter and shown to the user in a graph for the purpose of approximating the temperature or ionization parameter. The R ratio is commonly used as a diagnostic to indicate the density of the observed source, but assisting the user in approximating this value is beyond the scope of XSLIDE.

In the K diagnostics, ions for which K$\beta$ or K$\alpha$ lines have been identified in the spectrum will have the K$\beta$/K$\alpha$ ratio of the detected line fluxes calculated. This ratio is also commonly used as a diagnostic to indicate the temperature or ionization parameter of the observed source.

\subsection{Step 5: Export}

In this step, the results can be exported to various formats. This allows the results to be presented or utilized in other programs.

Output formats include a PDF report, PNG or SVG figures, and CSV or Excel tables of data. In addition, XSLIDE can export a model spectrum in a format that can be read directly into XSPEC\cite{Arnaud96}. This model spectrum will consist of the lines and continuum that have been detected in XSLIDE, giving the XSPEC user a starting point that can be further modified as desired.


\subsection{Other Features}

A key feature of XSLIDE is the interactive plot of the spectrum. The user can pan and zoom with intuitive controls to focus on particular features of the plot as desired. The user can also customize the plot by setting line colors, styles, widths, and symbols. In addition to being able to view the spectrum and the fitted continuum, the user can also choose to display the spectrum minus the continuum, the raw counts, and the ARF.

XSLIDE is available in both English and Japanese. The user can select the language as a menu option.

An extensive user manual is provided with XSLIDE which documents the various methods and parameters available in each step. It also provides a walkthrough of the software, demonstrating how XSLIDE can be used to approximately reproduce Figure 1 of Tashiro et al.\cite{Hitomi16}.

\section{Software Architecture}

XSLIDE has a Model-View-Controller (MVC) software design, which keeps a level of separation between the data and the GUI. “Data” refer to the physical quantities that XSLIDE is operating on, such as spectra, ARFs, continua, theoretical line lists, emissivity models, etc. The Model performs operations on the data elements, the View displays the current state of the Model to the user, and the Control part of the GUI manipulates data in response to user events by sending commands to operate on the Model.

This separation of programming elements allows for a central underlying Model to be displayed to multiple Views, allowing for separate desktop and web applications to be written with minimal reproduction of effort. It further simplifies development by providing a guided process by which to add a new feature, where first the Model is updated such that the new feature is available in the command line, then the View is updated such that new GUI elements showing the state of the Model are displayed, and finally the Control is updated so that the Model appropriately responds to user events. The separation helps with testing as well, as unit tests are written for the Model independently from functional tests written for the final GUI applications.

There are ultimately three XSLIDE products: a Python command-line-interface (XSLIDEpy), the desktop GUI (XSLIDE for desktop), and the web GUI (XSLIDE for web). The Model elements of the MVC are largely contained within XSLIDEpy, whereas the View and Controller elements are separately written for the desktop and web GUIs. Most users will only access either the desktop and/or web GUIs, with XSLIDEpy largely reserved for use in the development process.

XSLIDEpy is a Python package that leverages the standard scientific Python ecosystem of NumPy\cite{Harris2020array}, SciPy\cite{2020SciPy-NMeth}, and Astropy\cite{astropy:2013, astropy:2018} to quickly conduct common astronomical operations. XSLIDE uses PyVO, an Astropy-affiliated package, to query NASA's HEASARC\cite{heasarc} archive for spectrum and ARF files from the XRISM and Hitomi missions. Unit tests cover a large portion of the source code.

\begin{figure} [ht]
\begin{center}
\begin{tabular}{c}
\includegraphics[width=6cm]{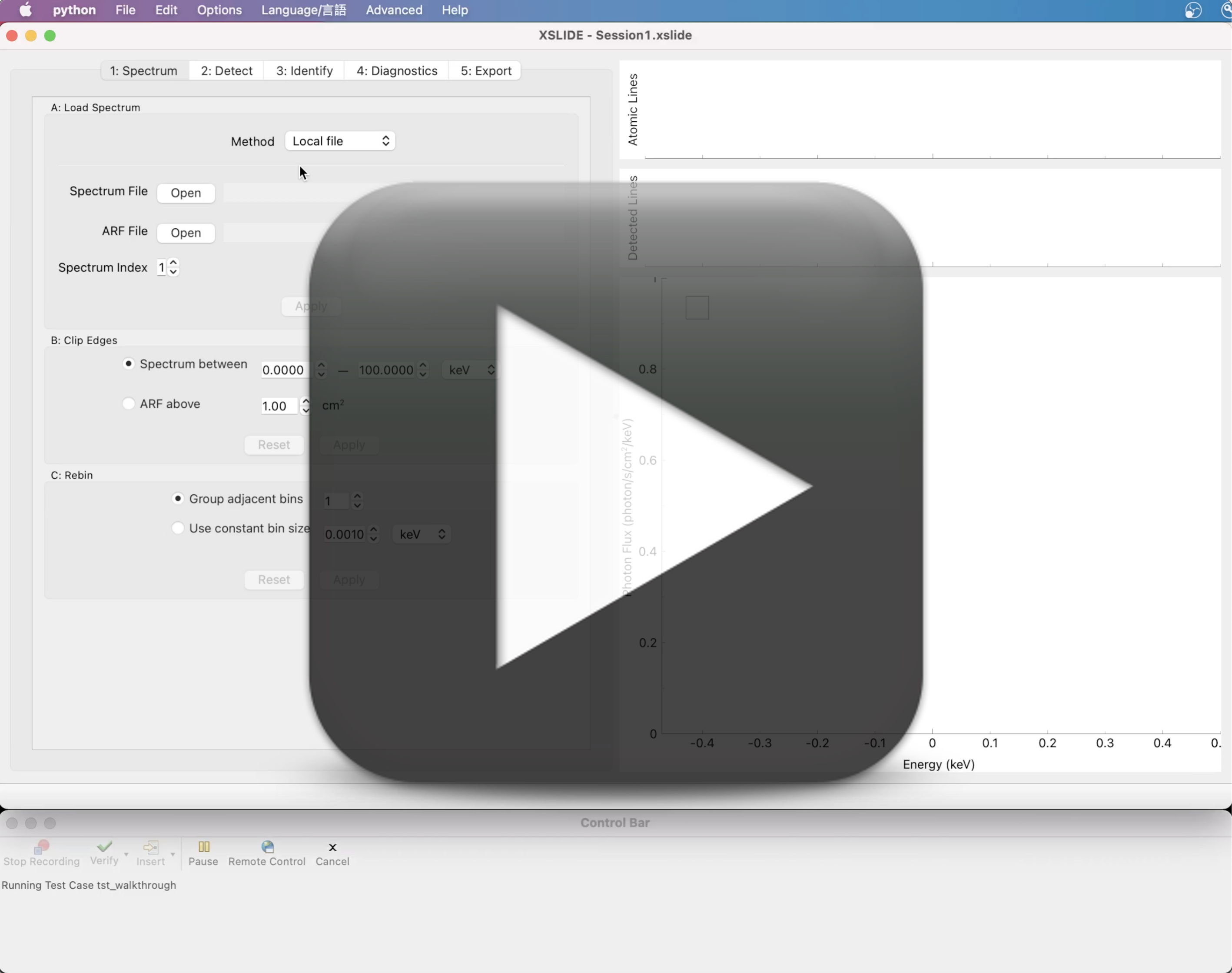}
\end{tabular}
\end{center}
\caption
{\label{fig:video1}
A video of Squish\cite{squish} performing one of XSLIDE's functional tests can be accessed as an ancillary file.}
\end{figure}

XSLIDE for desktop is built using Qt\cite{qt}, which allows XSLIDE to be run cross-platform on Mac and multiple Linux distributions. The Qt for Python (PySide6) bindings are used to develop the Control, and PyQtGraph\cite{pyqtgraph} is used for the interactive plot. PyInstaller\cite{pyinstaller} is used to bundle the application into a single package that can be installed via simple drag-and-click. Squish\cite{squish} is used to run automated functional tests that cover most of the functionality available to the end user. Squish runs the application through scripts of user actions, recognizing GUI objects and interacting with them by performing mouse and keyboard clicks, and then verifying that these actions produced the expected results by confirming that GUI objects contain predetermined values and that screenshots are visually equivalent to prepared standards. A video of Squish running one of the functional tests can be seen in the link provided in Fig.~\ref{fig:video1}.

XSLIDE for web is built using HTML, CSS, and Bokeh\cite{bokeh} to design the View and interactive plot. JavaScript and a Bokeh server are used to develop the Control. Selenium\cite{selenium} is used to run functional tests in a manner similar to what is done by Squish for the desktop version.

For both the desktop and web versions of XSLIDE, Japanese language translations are provided using Qt's QTranslator class and user documentation is prepared with Sphinx\cite{sphinx}.

\acknowledgments 

We thank Lorella Angelini for laying the foundational groundwork of XSLIDE, Tim Kallman for generating the XSTAR atomic database files, Megumi Shidatsu for assistance in creating Japanese translations, and Edmund Hodges-Kluck and Kenji Hamaguchi for providing invaluable user feedback that helped guide the development process.

\bibliography{main.bbl}

\end{document}